\documentstyle[11pt]{article}
 
\begin{document}
 
% \thesaurus{
%	03(12.03.3;12.12.1)	% Distances, distance scale,
%07.19.1;	% Galaxies: redshifts of,
%18.02.1;	% Radial velocities: {\sl see also Galaxy (the):kinematics and}
%20.01.2)	% Universe (the): structure of.
%	}
%
\begin{center}
 
{\bf One Arc Degree Core Substructure of the Virgo Cluster }
 
\vspace{0.2in}

% \author{
A.R.~Petrosian$^1$
%\inst{1} 
V.G.~Gurzadyan$^{2,3}$
%\inst{2,3} 
M.A.~Hendry$^{3,4}$
%\inst{2} 
and E.~Nikoghossian$^1$
%\inst{1}
%}	

\vspace{0.1in}
 
 %\institute{
1. Byurakan Astrophysical Observatory, Armenia 

2. University of Sussex, Brighton, UK

3. Department of Theoretical Physics, Yerevan Physics Institute,
Yerevan, Armenia (permanent address) 

4. University of Glasgow, Glasgow, UK
\vspace{0.2in}

\end{center}

 %\begin{abstract}

{\it Abstract}. The results of an analysis of the substructure  of 1 degree
field galaxies of the Virgo cluster by means of the S-tree technique
developed by Gurzadyan et al [28-31], are represented.
The existence of 3 main subgroups is shown and
their dynamical parameters are obtained. The mass centers of the
subgroups appear to be aligned in the direction parallel to
the elongation of the Virgo cluster. The morphological analysis
shows some domination of dwarf galaxies in the subgroup A containing M 87,
and existence of two spiral galaxies N4425 and N4461; the latter fact
can be crucial for
the estimation of the distance of the Virgo cluster by means of
the search of Cepheids in that spirals.

%\end{abstract}

\section{Introduction}

    The study of substructures in galaxy clusters is important
for the understanding of the evolution of  
galaxy clusters [1-3], as well as for cosmology [4-7].
Substructures seem to be often observed features in many clusters of
galaxies. Various studies (e.g. [8-12]) show that at least 30 - 50 per cent
of rich clusters reveal multi-component structure in galaxy distributions
and in X-ray images [13-16]. Evidence for the presence of
substructures in clusters of galaxies is anticipated also from theoretical
studies [17-19]. Some studies have shown that morphological segregation of
the galaxies in the cluster possibly has close connection to the cluster 
substructure [20,21].

   Virgo cluster is the nucleus of the Local Supercluster. It is an
irregular system. Virgo cluster has a core structure that is centered on M87
[22], and also is known by its complex substructure [23].
The large scale X-ray image of Virgo cluster [24]
is very similar to the structure in the galaxy distribution [25,26],

   Since Virgo cluster is the nearest large concentration of galaxies, 
it plays critical role in understanding  of the general properties of the
individual galaxies in the cluster and its relations with that
of the  whole system, and large-scale structure of the universe as well.
For example, the problem of orientation
of spin vectors of galaxies, which is an important tool to probe 
the epoch of the origin and formation
of galaxies and clusters, has been carried out for the Virgo cluster [27].

   In the  present paper we use S-tree method developed by Gurzadyan, 
Harutyunyan \& Kocharyan [28,29,31].
The method has been already applied for the study of the substructures
of the Local Group [32], and Abell [50,51] and Coma clusters [52]: e.g.
in the case
of the Local Group aside from confirmation of the general picture known
from other studies, some new associations between particular galaxies
have been observed [32]. The same geometrical approach can be also rather
informative while studying the fractal properties of the galaxy 
distribution [33,34].

The aim of the present study is to reveal the substructures in central -
around M87 (N 4486) - one degree field of Virgo cluster.
This study is the first step towards the revealing of the
Virgo cluster substructure by S-tree code. 
The number of galaxies in the catalog is being increased essentially with 
the increase of the field, so that the time required for dealing with the
phase space of many dimensional system by S-tree, increases sharply 
(nonlinearly), though, in principle, S-tree has no any constraint
on the limiting number of galaxies. 
We hope to perform the study of larger fields -- from 2 up to 
6 degrees, as well.  Because of the expected robust character of the main
conclusions of this preliminary study, as discussed in section 2, we decided
to represent them in the present paper.

       In section 2 we summarize the basic principles underlying
the S-tree technique. In section 3 we describe 
our subsample of Virgo galaxies and present the results of 
application of the S-tree method to determine the 
hierarchical substructure of this sample. Finally, 
in section 4, we discuss the obtained results, morphology of
galaxies, and also identify suitable candidate
galaxies for future Cepheid observational programs. This fact can be of
particular importance, given the crucial role of the Virgo cluster  
in calibration of the cosmological distance scale [35],
and especially, because of the measurements of the distance
of Virgo cluster via NGC 4571 and M100 [35,36] and some other galaxies,
aiming the determination of the Hubble constant.

\section{The S-Tree Technique}

Here we outline the main points of S-tree technique, 
referring for details to papers [28-30] and to the monograph [31].
 
S-tree is based on the property of structural stability, well known in 
the theory of dynamical systems, enabling to reveal   
the robust properties of the systems based on a limited amount of information.
Gravitating systems being 
exponentially unstable systems, therefore, are relevant for this purpose.

The main motivation of the developing of various new statistical methods here,
is the radical difference of the 
problem of dynamics of clusters of galaxies from the stellar dynamical
problem. This difference once more became clear while developing a
method of reconstruction of the transversal component of the
velocity of clusters of galaxies in [38].

Two main concepts are the basic ones for the S-tree technique:
the introduced {\it degree of boundness} of various members of the system 
(galaxies), and the corresponding {\it tree-diagram}
to representation of hierarchical substructure of the clusters.

The transition from the degree of correlation to so-called $\rho$-boundness  
is realized using the properties
of trajectories the systems in $6N$-dimensional phase space, namely by
their deviation on given value $\rho$ in a properly defined measure.
The problem of the dynamics of the N-body system is thus reduced to
the study of the geometrical properties of the phase space, 
which in accord to methods developed in theory of dynamical systems,
can be described via the two-dimensional curvature $K_{\mu \nu}$.
The degree of boundness is thus related to
the following matrix:
$$
D_{ab}= max_{i,j} \{ -K_{\mu \nu}, 0\};
$$
$$
\mu=(a,i), \nu=(b,j); 
a,b=1,2,..,N; i,j=1,2,3.
$$
where the tensor $K$ is determined via the Riemannian curvature 
$R^{\mu}_{\lambda\nu\rho}$ ($u$ is the velocity of geodesics)
$$
K^{\mu}_{\nu}=R^{\mu}_{\lambda\nu\rho}u^{\lambda}u^{\rho}
$$
of the Riemannian manifold defined by a metric depending on the potential of
interaction (see [39]).

The results of the calculations describing
the hierarchical substructure
of the system are represented by the tree-diagram ({\it S-tree})
via the corresponding transition to  matrices $\Gamma_{ab}$ 
from $D_{ab}$:
$$
\Gamma_{ab}(\rho)= \{ 0\, 1\}
$$
The input data for the S-tree analysis of a cluster of galaxies 
are as follows: the 2D coordinates of the galaxies, their
redshifts and the magnitudes. 

The latter values via the assumption $M=const L^n$, where $L$ denotes the
luminosity, $M$ - the mass of the galaxies, are used to involve the information
on the masses of the galaxies. The present calculations have been performed 
for $n=1$, though the results on the subgrouping and the membership 
of galaxies have been shown to be robust while moving from 
$n=1$ to $n=1/2$.

The results of the S-tree have been compared with
those of wavelet analysis, in particular, for clusters of ESO Key program
on nearby Abell clusters survey (ENACS). The parallel analysis of
the same cluster data by both methods has revealed the general coincidence  
of the results in defining of the main physical system of the cluster.
However, S-tree enables also to resolve the smaller
subsystems, in some cases removing the difficulties associated
with physically anomalous parameters of apparent substructures indicated
by the wavelet.

An essential point in the context of the present study is
in which degree the results of the substructure
analysis of the sample based on 1 degree field of the cluster can be
modified if the larger areas should be also involved. Numerous
studies, both of toy models, as well as of real clusters with artificially
cancelled fields have shown the following. Since the method is based on the
discovery of correlation between the members (galaxies) of the systems, 
any found correlation cannot disappear with adding of new areas, unless the
information on the previous members is changed drastically. Therefore
the added new areas supply new galaxies to the existing
subgroups or new correlations could appear as well, without
the distortion of the main subgrouping picture. Though some change in the
membership can occur, i.e. a galaxy from one subsystem can move to 
another one. It is natural, since some galaxies could be well attracted
almost equally by the two subgroups; typically the number of such galaxies
is not large (1-2), moreover they could be predicted by the analysis
of the initial (smaller) sample.
 
Therefore, the found substructure typically should be robust with respect 
to the increase of the field of consideration. Note only, that for system with
larger number of members (or of larger field), the normalization of
the degree of boundness is changed, since one deals with different
phase spaces. Practically this leads to the possibility of probing
more and more deeper structures (i.e. of more strongly correlated ones) with
the increase of the number of the members of the system.    

\section{Sample and Results}

   We applied the S-tree method to analyze a sample of galaxies lying
within 1 degree of NGC 4486, from the compilation of CfA catalogue [40].
Among the 125 listed galaxies within the 1
degree core of the cluster, there are 73 galaxies with measured redshifts,
and 66 with measured V-band
magnitudes. This sample of 66 galaxies is listed in Table 1, 
including the galaxy name, its
coordinates, apparent V-magnitude and heliocentric radial velocity.

\begin{table*}
\caption{Galaxies in 1 degree field with messured redshits and magnitudes - 66
objects}
\renewcommand{\tabcolsep}{4mm}
\begin{center}
\medskip
\begin{tabular}{lcccr}
\hline
\hline
 Galaxy    &  RA(1950)   &    Dec(1950) &    mv  &   V(km s-1) \\
\hline
1223+1226 &  12 23 56.9 &    12 26 25  &  16.60 &      16364  \\
N4413     &  12 24 00.0 &    12 53 00  &  13.04 &         96  \\
I3344     &  12 24 00.0 &    13 51 00  &  15.20 &       1375  \\
I3349     &  12 24 15.0 &    12 43 48  &  15.30 &       1471  \\
I3355     &  12 24 18.0 &    13 27 12  &  15.20 &         -9  \\
I3363     &  12 24 31.2 &    12 50 06  &  15.50 &        791  \\
N4425     &  12 24 42.0 &    13 00 42  &  13.21 &       1881  \\
N4431     &  12 24 55.2 &    12 34 06  &  14.50 &        913  \\
1225+1311 &  12 25 06.0 &    13 11 00  &  17.30 &      28322  \\
N4435     &  12 25 07.8 &    13 21 24  &  12.03 &        773  \\
N4436     &  12 25 10.2 &    12 35 30  &  14.80 &       1125  \\
N4438     &  12 25 13.6 &    13 17 07  &  12.00 &         86  \\
N4440     &  12 25 21.6 &    12 34 12  &  13.09 &        739  \\
1225+1215 &  12 25 22.7 &    12 15 56  &  17.40 &      26483  \\
1225+1157 &  12 25 29.9 &    11 57 25  &  17.50 &      20751  \\
1225+1221 &  12 25 34.2.&    12 21 18  &  15.85 &      26777  \\
I794      &  12 25 37.8 &    12 22 00  &  15.10 &       1934  \\
I3381     &  12 25 42.0 &    12 04 00  &  15.10 &        637  \\
1225+1324 &  12 25 42.0 &    13 24 00  &  17.60 &      28636  \\
I3388     &  12 25 55.8 &    13 05 54  &  15.40 &       1761  \\
N4452     &  12 26 12.0 &    12 02 00  &  13.33 &        223  \\
I3393     &  12 26 12.0 &    13 11 00  &  15.10 &        466  \\
N4458     &  12 26 25.8 &    13 31 06  &  13.32 &        684  \\
N4461     &  12 26 31.2 &    13 27 42  &  12.37 &       1925  \\
1226+1243 &  12 26 51.6 &    12 43 36  &  15.56 &        538  \\
1227+1234 &  12 27 06.0 &    12 34 00  &  17.50 &      26041  \\
N4473     &  12 27 16.8 &    13 42 24  &  11.61 &       2236  \\
1227+1330 &  12 27 18.0 &    13 30 19  &  17.40 &      13100  \\
1227+1157 &  12 27 20.3 &    11 57 13  &  16.00 &      25085  \\
1227+1218 &  12 27 39.3 &    12 18 51  &  16.80 &      16860  \\
1227+1346 &  12 27 45.9 &    13 46 09  &  17.10 &      24879  \\
N4478     &  12 27 46.2 &    12 36 18  &  12.57 &       1370  \\
N4479     &  12 27 46.8 &    13 51 12  &  13.93 &        858  \\
\hline
\end{tabular}
\end{center}
\end{table*}

\setcounter{table}{0}
\begin{table*}
\caption{Cont.}
\medskip
\renewcommand{\tabcolsep}{4mm}
\begin{center}
\begin{tabular}{lcccr}
\hline
\hline
 Galaxy    &  RA(1950)   &    Dec(1950) &    mv  &   V(km s-1) \\
\hline
1227+1244 &  12 27 49.0 &    12 44 22  &  18.50 &      26000  \\
N4486B    &  12 28 00.0 &    12 46 00  &  14.50 &       1586  \\
1228+1242 &  12 28 09.7 &    12 42 09  &  21.96 &      16285  \\
1228+1238 &  12 28 10.5 &    12 38 04  &  16.80 &      26285  \\
1228+1238 &  12 28 13.3 &    12 38 32  &  16.60 &      25396  \\
1228+1241 &  12 28 13.4 &    12 41 43  &  21.61 &      22430  \\
1228+1237 &  12 28 14.9 &    12 37 27  &  21.65 &       6310  \\
1228+1236 &  12 28 15.7 &    12 36 13  &  21.71 &      14000  \\
1228+1244 &  12 28 15.8 &    12 44 06  &  21.24 &      13160  \\
1228+1241 &  12 28 15.9 &    12 41 25  &  20.88 &       5150  \\
1228+1219 &  12 28 17.4 &    12 19 18  &  17.00 &       1250  \\
N4486     &  12 28 17.6 &    12 40 01  &  10.30 &       1292  \\
1228+1244 &  12 28 18.0 &    12 44 14  &  21.11 &       7000  \\
1228+1242 &  12 28 19.8 &    12 42 53  &  21.24 &       2925  \\
N4486A    &  12 28 24.0 &    12 33 00  &  11.20 &        450  \\
1228+1238 &  12 28 24.5 &    12 38 58  &  20.79 &       2735  \\
1228+1238A&  12 28 28.5 &    12 38 21  &  20.73 &      80320  \\
1228+1246 &  12 28 37.0 &    12 46 10  &  20.86 &      11095  \\
I3443     &  12 28 44.4 &    12 36 30  &  15.60 &       1814  \\
I3457     &  12 29 19.6 &    12 55 57  &  15.40 &       1469  \\
I3459     &  12 29 22.8 &    12 27 00  &  15.50 &        278  \\
I3466     &  12 29 33.0 &    12 05 36  &  15.30 &        786  \\
N4506     &  12 29 42.0 &    13 42 00  &  14.20 &        681  \\
I3461     &  12 29 51.0 &    12 10 12  &  15.50 &       1110  \\
I3467     &  12 29 52.2 &    12 03 48  &  15.40 &       7519  \\
1229+1204 &  12 29 54.0 &    12 04 00  &  15.40 &       7810  \\
I3475     &  12 30 04.8 &    13 03 00  &  14.94 &       2572  \\
I3489     &  12 30 42.0 &    12 31 00  &  15.20 &       7834  \\
I3492     &  12 30 42.0 &    13 08 00  &  15.60 &       2004  \\
I3492A    &  12 30 48.0 &    13 08 00  &  15.30 &       -571  \\
I3501     &  12 31 18.0 &    13 36 00  &  15.00 &       1608  \\
N4531     &  12 31 42.0 &    13 21 00  &  13.30 &          8  \\
I3509     &  12 31 48.0 &    12 21 00  &  15.30 &       2073  \\
\hline
\end{tabular}
\end{center}
\end{table*}

As is mentioned above, the S-tree analysis was performed for various
$M/L$ relations, but the results were stable.
Therefore here the results for $n=1$, i.e. with constant mass
to luminosity ratio, are shown only. The 
hierarchical substructure identified by the S-tree analysis indicate
that the main group contains 36 galaxies. Remaining
30 galaxies either are chance projections
and/or  background objects. This group itself consists of three 
subgroups - denoted A, B, and C
and containing 17, 12 and 6 galaxies, respectively. From 17 galaxies
of the M 87 subsystem (A), 10 are more strongly bounded
than the rest 7. Group A is the group containing M87, which
one should then most reasonably associate 
with the core of the Virgo cluster. The galaxy NGC 4473 does 
not belong to any of determined groups in the central 1 degree region,
though it contributes into the gravitational potential of
the whole central region of Virgo cluster.
In general, the S-tree results indicate that we deal with 
fractions of more larger subgroups, especially, in the case of groups
$B$ and $C$, so that the increasing of the field most probably would supply 
additional galaxies into these groups.

   In Figure 1 the projected
positions of the 36 galaxies are plotted, indicating via different symbols
the galaxies belonging to $A$, $B$ and $C$ 
groups. We see that not only the projected positions of these
galaxies but also the redshift information only is not enough to specify
the subgroups due to some overlapping of their redshift distributions
(Figure 2).

   Table 2 represents the list of galaxies, members of
A, B and C groups within the central 1 degree field of
Virgo cluster. In the Table 2 the following information is included: (1) the
name of the subgroups: (2) the name of
the galaxies, including their relative degree of boundness,
{\it s}-strong, {\it w}-weak,  and their 
radial velocities. For each galaxy the  B magnitude, 
diameter and axial ratio from RC3 [41],
the morphology is given according to [42]
and position angles (P.A.) from UGC [43].
For those galaxies which are not included in
UGC, P.A. are measured by ourselves on red prints of POSS.

\begin{table*}
\caption{Members of A, B and C Groups Within the Central 1 Degree
Field of the Virgo Cluster}
{\small
\medskip
\renewcommand{\tabcolsep}{6mm}
\begin{center}
\begin{tabular}{llrlr}
\hline
\hline
Group & Galaxy      &   B(T)  &    Morph   &  PA  \\
\hline
Aw    & N4425       &  12.82  &    SBa     &  27  \\
As    & I3344       &  14.80  &    dE6     &  54  \\
As    & I3349       &  14.78  &    dE1     &  43  \\
Aw    & I3388       &  15.31  &    dE5     & 101  \\
Aw    & I794        &  14.73  &    dE3     & 110  \\
As    & N4436       &  14.03  &    dE6/dS  & 118  \\
Aw    & N4461       &  12.09  &    Sa      &   9  \\
As    & N4478       &  12.15  &    E2      & 140  \\
As    & N4486       &   9.58  &    E0      &   -  \\
As    & 1228+1219   &  16.80  &    BCD     & 152  \\
As    & N4486B      &  15.11  &    E1      &  63  \\
Aw    & I3443       &  15.64  &    dE0     &   -  \\
As    & I3457       &  15.40  &    dE3     &  47  \\
As    & I3461       &  14.82  &    dE2     &  30  \\
Aw    & 3492        &  14.58  &    E3/S0   & 116  \\
Aw    & I3509       &  14.73  &    E4      & 121  \\
As    & I3501       &  14.50  &    dE1     &   -  \\
B     & N4413       &  13.97  &    SBbc    &  60  \\
B     & I3355       &  14.82  &    SBm     & 172  \\
B     & N4438       &  10.90  &    SB      &  27  \\
B     & N4452       &  13.30  &    S0      &  32  \\
B     & I3459       &  14.62  &    dSB0    &   -  \\
B     & N4531       &  12.58  &    Sa pec  & 155  \\
C     & I3363       &  15.40  &    dE7     & 127  \\
C     & N4431       &  13.72  &    dS0     & 177  \\
C     & N4435       &  11.84  &    SB0     &  13  \\
C     & N4440       &  12.74  &    SBa     &  94  \\
C     & I3381       &  14.95  &    dE4     & 110  \\
C     & 1226+1243   &  15.56  &    dE1     &  14  \\
C     & N4458       &  12.92  &     E1     &  46  \\
C     & I3393       &  14.60  &    dE7     &  47  \\
C     & N4479       &  13.45  &    SB0     &   -  \\
C     & N4486A      &  13.40  &     E2     & 170  \\
C     & I3466       &  15.72  & pec, N     &  29  \\
C     & N4506       &  13.64  &  S pec     & 110  \\
MS    & N4473       &  11.10  &     E5     & 100  \\
\hline
\end{tabular}
\end{center}}
\end{table*}

\section{Discussion}

\subsection{The membership of galaxies in groups}

The galaxies of our sample lie in the region of Virgo cluster [44]. which
has average heliocentric velocity 991 km/s and velocity dispersion 661
km/s. The dynamical characteristics of the Virgo 1d field subgroups A, B and
C are given
in Table 3, including the average redshifts (V) and the velocity dispersions
($\sigma$); the results for the 17 galaxies of subgroup A and for its core
of 10 galaxies are given separately.
\begin{table*}
\caption{}
\medskip
\begin{tabular}{llllll}
\hline
\hline
Subgroup & N   & V (km/s)& $\sigma$(km/s)& \\
\hline
A (core)&  10  & 1366    & 174    & \\
A       &  17  & 1473    & 480    & \\
B       &   6  &  113    &  115   & \\
C       &  12  & 693     &  148   & \\
\hline
\end{tabular}
\end{table*}

\subsection{Spatial distribution of substructures}
Already the visual inspection
of  Figure 2 gives an
impression that A, B and C subgroups of galaxies in the center of
Virgo cluster have collinear distribution.
In order to quantify this impression, we have determined the 
coordinates of their mass centers:
$$
     A :\,    \alpha_1  = 12^h 28^m 55.83^s;\,    \delta_1 = 12^d 30^m 16.196^s
$$
$$
     B :\,    \alpha_2 = 12^h 25^m 31.00^s;\,     \delta_2 = 12^d 32^m 53.710^s
$$
$$
     C :\,    \alpha_3 = 12^h 27^m 15.61^s;\,     \delta_3  = 12^d 32^m 50.110^s
$$
  To test their alignment we use the equation:
$$
   \tan (\delta_1) \sin(\alpha_3 - \alpha_2) + \tan (\delta_2) \
\sin(\alpha_3 - \alpha_1) + \tan (\delta_3) \sin(\alpha_1 - \alpha_2)=0,
$$
i.e. the equation of the three points $(\alpha_i, \beta_i)$ situated 
on a line. Our computations yield for the left hand side $0.01$, i.e. the
mass centers of $A$, $B$ and  $C$ groups are well aligned.  

   The collinear distribution of the substructures in the
centre of the Virgo cluster  is not a unique one. 
According to [16] there is a tendency for alignment in the substructures
in clusters. The position angle of the line connecting A, B and C subgroups
is $92^{\circ}$. According to [45] the P.A. angle of
the jet of M 87 is $290^{\circ}.80\pm0.5$, while the P.A. of
major axis of the Virgo S cluster is about $100^{\circ}$.

The revealed three subgroups also show some elongation. In the Table 4 we
represent
the P.A. of the major axes of A, B and C, the projected distance
of their major axes from their mass centers (d) and the major diameter
of each subgroup.
\begin{table*}
\caption{}
\medskip
\begin{tabular}{llllll}
\hline
\hline
Subgroup & P.A.& d (kpc)&   D(kpc)& \\
\hline
A (core)&    98  & 90   & 1200   & \\
A       &    92  &  6   &   80   & \\
B       &   172  & 63   &  600   & \\
\hline
\end{tabular}
\end{table*}
Note that, the major axes of A and B are almost parallel to the direction
of alignment of subgroups, while the major axes of C is almost
perpendicular to it.

We had looked also for the angular distribution of the galaxies within each
subgroup. The formula [45]
$$
N(\theta)= \mu [1+\Delta_1\cos(2\theta_i)+\Delta_2\sin(2\theta_i)],
$$
where
$$
\Delta_1=[\Sigma N(\theta)\cos(2\theta_i)]/3\mu;
\Delta_2=[\Sigma N(\theta)\sin(2\theta_i)]/3\mu;
$$
has been used; $N(\theta_i)$ is the number of galaxies
with position angle in i-th binning, $\mu$ is the mean surface density
of galaxies. This formula
yields $\Delta_{1,2}=0.1-0.9$, thus indicating no any significant
departure from the homogeneous distribution. This is not surprising
given the small number of galaxies in the samples.

\subsection{Morphology of Galaxies in Subgroups}
    In the Table 5 data for galaxies morphological
distribution  in the A, B and C groups are presented.   
Morphological types are sampled as ellipticals (E), spirals (S), 
lenticulars (S0) and dwarfs (dE,
dS0 and BCD galaxies). Some remarkable features can be
noticed. One concerns, for example, the population of dwarf galaxies in the
subgroup A; seven from ten of those galaxies are located in projection
within 90 degrees with respect to the direction of the jet of
M 87. The same subgroup contains also two spirals N4425 and N4461. Subgroup B
is dominated with spirals, with no ellipticals.
Dwarf galaxies are the majority also in group C.
\begin{table*}
\caption{}
\medskip
\begin{tabular}{llllll}
\hline
\hline
Subgroup & E   &  S0     &  S &  Dwarfs & \\
\hline
A       &   5  &         &  2    &  10  & \\   
A (core)&   3  &         &  2    &   2  & \\
A       &   3  &         &       &   7  & \\
B       &      &   1     &  4    &   1  & \\
C       &   2  &   2     &  2    &   6  & \\
\hline
\end{tabular}
\end{table*}
Though one should be cautious in drawing any far going conclusions from these
morphological distributions, and moreover, these results have no reason to
coincide with the populations in clusters, nevertheless, note, that the
E/SO/S ratio is different than what is known for clusters of various richness
[20].

\subsection{Cepheids and the Virgo Distance Problem}

The existence of the two spiral galaxies in the subgroups A, i.e. the one
including the M 87, revealed by the present study can provide
possibility for the estimation of their distances by means of the
search of Cepheids in those spirals. This fact can be crucial for
the estimation of the Hubble constant [46-49].

\section{Conclusions}

The present study enables us to draw the following
main conclusions:

(a) Three main subgroups are revealed by means of the study of 1 arc degree
central field of the Virgo cluster;

(b) The dynamical parameters of the 3 subgroups are estimated.
Alignment of the mass centers of the subgroups correlates
with the elongation of the Virgo cluster.

(3) The subgroups themselves show some elongation in projection, so that
the elongation of the  subgroups A and B is parallel to the
the alignment of the subgroups, and is perpendicular
to that of the subgroup C.

(4) The subgroup A is dominated by dwarf galaxies,
with some preference in their location  in the direction of the jet of M87.
No ellipticals exist in the subgroup B.

(5) The presence of two spirals N4425 and N4461 in the subgroup A
can provide possibility for the estimation of the distance to the
core of the Virgo cluster via Cepheids.

The studies of the larger fields of the Virgo cluster will be
desirable for the evaluation of the physical content of these
conclusions, as well as for the general aim of the understanding of the role
of subgroups in the dynamics of clusters of galaxies [50-52].

Authors are thankful to A.Melkonian for assistance with computations.
V.G. is grateful to J.Barrow
for the hospitality in Sussex Astronomy Centre where he was supported
by the Royal Society.

\vspace{0.2in}

REFERENCES

\begin{description}

\item[1.] West M.J., Oemler A.,Dekel A. ApJ 327, 1, 1988     \\[-6mm]
\item[2.] Beers T.C., Geller M.J.  ApJ, 274, 491, 1983       \\[-6mm]
\item[3.] Tremaine S.,  in Dynamics and Interactions of Galaxies, ed.
    R. Wielen, Springer, N.Y., 394, 1990 \\[-6mm]
\item[4.] Richstone D., Loeb A., Turner E.L.  ApJ, 393, 477, 1992\\[-6mm]
\item[5.] Kauffmann G., White S.D.M. MNRAS, 261, 921, 1993       \\[-6mm]
\item[6.] Lacey G.G., Cole S. MNRAS, 262, 627, 1993              \\[-6mm]
\item[7.] Jing Y.P., Mo H.J., Borner G., Fang L.Z. MNRAS, 276, 417, 1995 \\[-6mm]
\item[8.] Geller M.J., Beers T.C. PASP, 94, 421, 1982      \\[-6mm]
\item[9.] Bird C. AJ, 107, 1637, 1994                      \\[-6mm]
\item[10.] Escalera E., Biviano A., Girardi M., Giuricin G.,
    Mardirossian F., Mazure A., Mezzetti M.  ApJ, 423, 539, 1994 \\[-6mm]
\item[11.] Maccagni B., Garilli M., Tarenghi M. AJ, 109, 465, 1995 \\[-6mm]
\item[12.]  West M.J. in: Clusters of Galaxies, Eds.F.Durret, A.Mazure and
    J.Tran Thahn Van, Editions Frontieres, 1994                    \\[-6mm]
\item[13.] Mohr J.J., Fabricant D.G., Geller M. ApJ, 413, 492, 1993 \\[-6mm]
\item[14.] Grebenev S.A., et al  ApJ, 445, 607, 1995                \\[-6mm]
\item[15.] Zabludoff A.J., Zaritsky D. ApJ 447, L21, 1995           \\[-6mm]
\item[16.] West M.J., Jones Ch., Forman W.  ApJ, 451, L5, 1995      \\[-6mm]
\item[17.] White S.D.M.  MNRAS, 177, 717, 1976                      \\[-6mm]
\item[18.] Gonzalez-Casado G., Mamon G.A., Salvadore-Sole E. in: Clusters
    of Galaxies, Eds.F.Durret, A.Mazure and J.Tran Thahn Van,
    Editions Frontieres, 1994                                   \\[-6mm]
\item[19.] Cavaliere A., Colafrancesco S., Menci N. in: Cluster and
    Superclusters of Galaxies, ed. A.C. Fabian, Dordrecht, Kluwer, 331,
    1992                                                        \\[-6mm]
\item[20.] Caon N., Einasto M. MNRAS, 273, 913, 1995            \\[-6mm]
\item[21.] Doi M., Fukugita M., Okamura S., Turner E.L.  AJ, 109, 1490,
    1995                                                    \\[-6mm]
\item[22.] Nulsen  P.E.J., Bohringer H.  MNRAS, 274, 1093 , 1995 \\[-6mm]
\item[23.] Springel V. et al MNRAS 1997 (submitted); astro-ph/9710368 \\[-6mm]
\item[24.] Bohringer H., Briel U.G., Schwarz R.A., Voges W., Hartner G., Trumper J.
    Nature, 368, 828, 1994                                            \\[-6mm]
\item[25.] Binggeli B.,  Tammann G.A., Sandage A.  AJ, 94, 251, 1987  \\[-6mm]
\item[26.] Binggeli B.,  Popescu C.C., Tammann G.A. A\&AS, 98, 275, 1993 \\[-6mm]
\item[27.] Hu F.X., Wu G.X., Su H.J., Liu Y.Z. A\&A, 302, 45, 1995    \\[-6mm]
\item[28.] Gurzadyan V.G., Harutyunyan V.V., Kocharyan A.A. in: Proc. II
    DAEC Meeting, Eds. G.Mamon, D.Gerbal, Obs. de Paris Publ, 333, 1991 \\[-6mm]
\item[29.] Gurzadyan V.G., Harutyunyan V.V., Kocharyan A.A. A\&A, 281 964,
  1994                                                           \\[-6mm]
\item[30.] Bekarian K.M., Melkonian A.A. Astrofizika, 40, 425, 1997;
    astro-ph/9704054.                                              \\[-6mm]
\item[31.] Gurzadyan V.G., Kocharyan A.A. '{\it
    Paradigms of the Large-Scale Universe}',
    Gordon and Breach, N.Y., 1994                                  \\[-6mm]
\item[32.] Gurzadyan V.G., Kocharyan A.A., Petrosian A.R. Ap Space Sci.
    201 243, 1993                                                  \\[-6mm]
\item[33.] Gurzadyan V.G., Kocharyan A.A. Europhys.Lett., 15, 801, 1991 \\[-6mm]
\item[34.] Monaco P. A\&A, 287, L13, 1994                          \\[-6mm]
\item[35.] Rowan-Robinson M. in: Observational Tests of
    Cosmological Inflation, ed H.T. Shanks, Dordrecht, Kluwer, 1991 \\[-6mm]
\item[36.] Pierce M.J. et al Nature, 371, 385, 1994                 \\[-6mm]
\item[37.] Freedman W.L. et al  Nature, 371, 757, 1994              \\[-6mm]
\item[38.] Gurzadyan V.G., Rauzy S. Astrofizika, 40, 473, 1997;
    astro-ph/9707198                                                \\[-6mm]
\item[39.] Arnold V.I. {\it Mathematical Methods in Classical Mechanics}, Nauka,
    Moscow, 1989.                                                   \\[-6mm]
\item[40.] Huchra J.P. et al CFA Redshift Catalogue, Harvard-Smithsonian
    Center for Astrophysics, Cambridge, 1993                        \\[-6mm]
\item[41.] de Vaucouleurs G., de Vaucouleurs A., Corwin H.G., et al.
    Third Reference Catalogue of Bright Galaxies, Springer, Berlin (RC3),
    1991                                                            \\[-6mm]
\item[42.] Binggeli B., Cameron L.M.  A\&AS, 98, 297, 1993          \\[-6mm]
\item[43.] Nilson P.  Uppsala General Catalogue of Galaxies,
    Nova Acta Uppsala University, Ser. V:A,                         
    Vol.1 (UGC), 1973                                               \\[-6mm]
\item[44.] Nolthenius R., ApJS, 85, 1, 1993                         \\[-6mm]
\item[45.] de Vaucouleurs G., Nieto J-L. ApJ, 231, 364, 1979        \\[-6mm]
\item[46.] Tammann G.A. et al in: Science with the Hubble Space Telescope -
    II, Eds.P.Benvenuti, F.D.Maccheto, E.J.Schreier,  Space Telescope
    Institute, 1996                                                 \\[-6mm]
\item[47.] Federspiel M., Tammann G.A., Sandage A. ApJ 1997 (in press);
    astro-ph/970940                                                 \\[-6mm]
\item[48.] Bohringer H. et al ApJ 1997 (in press); astro-ph/9703012  \\[-6mm]
\item[49.] Gonzalez A.H., Faber S.M. ApJ 1997 (in press);
   astro-ph/9704192   \\[-6mm]
\item[50.] Gurzadyan V.G., Mazure A. The Observatory, 116, 391, 1996  \\[-6mm]
\item[51.] Gurzadyan V.G., Mazure A. MNRAS 1997 (in press);
   astro-ph/9709210  \\[-6mm]
\item[52.] Gurzadyan V.G., Mazure A. in: Proc. 'Coma Cluster' workshop,
(Marseille, 1997) (in press); astro-ph/9710094  \\[-6mm]

\end{description}

Figure Captions.
\vspace{0.1in}

{\it Figure 1}.
The distribution of galaxies in 1 arc degree field of Virgo cluster.
Triangles denote the galaxies of subgroup A, crosses - of
subgroup B, diamonds - of subgroup C.

{\it Figure 2.}
The redshift histrograms of the subgroups; the notations are the same
as in Figure 1.

\end{document}